\documentclass[12pt]{article}
\usepackage{graphicx}
\usepackage{amssymb,amsmath,amsfonts,palatino,amsthm}
\usepackage{amssymb}
\usepackage{epstopdf}
\newcommand{\beq}{\begin{equation}}
\newcommand{\eeq}{\end{equation}}

\newcommand{\f}{\begin{equation}}
\newcommand{\ff}{\end{equation}}

\begin{document}

\title{The Plebanski action extended to a unification of gravity and Yang-Mills theory }
\author{
Lee Smolin\thanks{Email address:
lsmolin@perimeterinstitute.ca}\\
\\
\\
Perimeter Institute for Theoretical Physics,\\
31 Caroline Street North, Waterloo, Ontario N2J 2Y5, Canada}
\date{\today}
\maketitle
\vfill
\begin{abstract}
We study a unification of gravity with Yang-Mills fields based on a simple extension of the
Plebanski action to a Lie group $G$ which contains the local lorentz group.  The Coleman-Mandula theorem is avoided because the dynamics has  no global spacetime symmetry. 
This may be applied to Lisi's proposal of an $E8$ unified theory, giving  a fully $E8$ invariant action.   The extended form of the Plebanski action suggests a new class of spin foam models.

\end{abstract}
\tableofcontents
\vfill
{\it This paper is dedicated to the memory of my teacher Sidney Coleman.  } 
\vfill
\newpage

\section{Introduction}

Ashtekar's formulation of general relativity\cite{abhay} taught us to think of gravitational theories as theories of connections, on a bare manifold
with no metric structure. This insight is deepened by the Plebanski form of the action\cite{plebanski} and related forms  studied by Capovilla, Dell and Jacobson and others\cite{CDJ,robertolaurent} which reveal that general relativity is a simple perturbation or constraint of a topological field theory.   This taught us, for example, that the metric is not fundamental, it emerges from solutions to the classical field equations.  These insights of the structure of the classical theory led to much progress on the quantization of gravity, in loop quantum gravity and spin foam models\cite{carlobook}. 

The idea that general relativity has its deepest formulation as a connection theory suggested immediately a new approach to the unification of general relativity with Yang-Mills theories.  The group of the connection in the connection formulations of general relativity is the local lorentz group $SO(3,1)$ or a subgroup of it.  What if one just takes a larger group $G$ containing $SO(3,1)$ and plugs it into the same action or hamiltonian constraint that gives general relativity so that one now has a theory with local $G$ invariance?  Does one get a theory that contains general relativity coupled to Yang-Mills fields in a subgroup of $G/SO(3,1)$?  

This question was answered affirmatively, in the Hamiltonian formulation, by Peldan\cite{peldan}.  All that was required was a simple modification of the  hamiltonian constraint, to keep the field equations consistent.   Peldan's approach was further studied by  Peldan and Chakraborty\cite{pc}, and Gambini, Olson and 
Pullin\cite{gp}\footnote{Other, possibly related approaches to unification of the gauge groups of general relativity and Yang-Mills theory are explored  in \cite{related}.}

More recently, it has been understood by Krasnov\cite{krasnov} and  Bengtssom\cite{ben} that these extensions of general relativity have elegant formulations in terms of action principles which  extend the Plebanski action in a natural way.  In this paper we study a very simple theory of this kind.  
We find that the addition of one simple term to the Plebanski action suffices to get a consistent dynamics for any $G$ containing the local Lorentz group.  There is a simple mechanism which breaks the symmetry down in such a way that the resulting dynamics is Yang-Mills  coupled to general relativity plus corrections. The latter  involve gauge fields which mix local lorentz transformations with local internal gauge transformations.  

One reason for revived interest in these kinds of unifications is Lisi's recent proposal\cite{lisi} of a unified theory, based on the same strategy, where $G=E8$.  There are several open issues regarding this proposal, which would have to be resolved for the idea to rise to the level of a theory.  At least two of them
have to do with making the dynamics fully $E8$ invariant.    Lisi proposes a form of the action based  on an approach to writing general relativity as a gauge theory, invented by MacDowell and 
Mansouri\cite{mm}.  In that scheme the metric and local Lorentz connection are unified in a deSitter or AdS connection, but the action is only invariant under
a local Lorentz subgroup.   Lisi's action is similar, with $E8$ playing the role of the deSitter group.   The action studied here, 
applied to $E8$, gives a fully gauge invariant action, which has solutions which spontaneously break the symmetry and give, when expanded around,  the bosonic part of Lisi's action, plus corrections.  

Another way in which Lisi's proposal breaks the gauge invariance is by a strategy of  incorporate fermions  by means a $BRST$ extension of the connection.      In section 3.1 below I propose an alternative way to incorporate the fermions, which would not break the gauge symmetry.  It is based on proposals that matter degrees of freedom arise in loop quantum gravity as  a result of the phenomena of disordered 
locality discovered by Markopoulou\cite{disordered,disordered2}.  This is in fact a version of a proposal of  Misner and Wheeler from 1957 that matter might be nothing but the mouths of Planck scale wormholes\cite{Wheeler} and, in   the context of loop quantum gravity, this was argued previously to give fermions\cite{me-fermions}.  In this kind of mechanism, the  full local symmetry remains present, but the cost is that for every generator of  $G$ there is both a gauge boson and a fermion degree of freedom.  

Regarding the quantization, the class of actions proposed here suggest a particularly simple class of spin foam models.  This is discussed in section 3.2 below.  

There are many aspects of the theories proposed here to be developed.  In general, and in regard to $E8$, there are many open issues and many things to work out.  For example in the latter there are open issues concerning  how the structure of the generations is realized. Nonetheless, these approaches are worth exploring because it is very possible that they give rise to consistent, finite quantum theories.  Many of the kinematical results of loop quantum gravity will go over to them, and as for dynamics,  they involve simple modifications of by now well studied spin foam models.   

In the next section we describe the general extension of the Plebanski action and show how it reduces to Einstein-Yang-Mills plus corrections.  An interesting example of this kind of unification is illustrated by the case of $SO(7,1)$ (or $SO(8)$ in the Euclideanized theory.)  This yields a unification of gravity and the electroweak theory, of the kind studied before in \cite{graviweak}, and includes naturally the Higgs bosons.  This is a sector of Lisi's proposed $E8$ unification, and will be studied in a separate paper\cite{SGL}

We close the introduction by noting that the well-known Coleman-Mandula no-go theorem\cite{cm} is avoided because that only applies to an $S$-Matrix whose symmetries include global Poincare invariance.  This theory, like general relativity, has no global symmetries, the Poincare symmetry acts only on the ground state not the action, and only in the limit in which the cosmological constant is zero. In fact, there is a nonzero cosmological constant,  as it is related to parameters of the theory.  By the time the $S$ matrix in Minkowski spacetime could be defined in this theory one will be studying only small perturbations of a ground state in a certain limit and the symmetry will only apply in that limit and approximation. As we shall see below, the symmetry will already be broken by the time that approximation and limit are defined, in such a way that Coleman-Mandula theorem could be satisfied in its domain of applicability.

\section{Plebanski action for a general group}

We will study here a general extension of the Plebanski action based on a connection valued in a semisimple $G$ which contains the Euclidean Lorentz group $SO(4)$.  (For simplicity we consider only the case of  Euclidean signature.)

We begin by reviewing the Plebanski action for general relativity, based on the 
group $SO(4)$,  parameterized by antisymmetric pairs of four dimensional vector indices, $[ab]$,
where $a=1,...4$, 
\f
S^{Pleb}=\frac{1}{G} \int_{\cal M} B^{ab} \wedge F_{ab} -\frac{1}{2}\phi_{abcd} B^{ab} \wedge B^{cd}
\label{action0}
\ff
The integral is over a four manifold, $\cal M$.  $F^{ab}$ is the curvature of the spacetime connection $A^{ab}$ that gauges the local Lorentz symmetry, $B^{ab}$ is a two form valued in the Lie algebra of $SO(4)$ and $\phi_{abcd}$ is a multiplet of scalar
fields that satisfy,
\f
\phi_{abcd}= \phi_{cdab}=-\phi_{bacd}
\ff
We do the variation subject to the constraint that
\f
\epsilon^{abcd}\phi_{abcd}= \Lambda, 
\ff
where $\Lambda$ the cosmological constant.  For the derivation of the Einstein field equations from (\ref{action0}), see (\cite{robertolaurent}), how this works will also be clear shortly.

We now extend this by embedding $SO(4)$ in a larger Lie algebra, $G$ and adding one term to the action.  The new, extended,  Plebanski action is (generators of
$G$ are labeled by $I,J=1,...,n$.)
\f
S^{G}=\frac{1}{G} \int_{\cal M} \left [ B^J \wedge F_J -\frac{1}{2}\phi_{JK} B^J \wedge B^K
+\frac{g}{2} \phi^{KL} \phi_{KL} B^J \wedge B_J
\right ] 
\label{action1}
\ff
Now, $A^I$ is a $G$-connection, $B^I$ is a two-form valued in the lie algebra of $G$ and $\phi_{IJ}=\phi_{JI}$ is a multiplet of scalar fields.   $A$ has dimensions of inverse length, $B$ is dimensionless so $\phi$ has dimensions of $(length)^{-2}$.   $g$ is a new coupling constant with dimensions of $length^2$.

The field equations are,
\f
F_J= \phi_{JK} B^K - g \phi^{KL} \phi_{KL} B^J
\label{eq1}
\ff
\f
{\cal D} \wedge B^K =0
\label{eq2}
\ff
\f
B^{(J} \wedge B^{K)} = 2g  \phi^{JK}B^L \wedge B_L
\label{eq3}
\ff

Let us now consider the consequences of the last field equation, (\ref{eq3}), which resulted from varying the 
$\phi^{IK}$

Tracing  (\ref{eq3}) we find, so long as $B^L\wedge B_L \neq 0$, 
\f
\phi^K_K = \frac{1}{2g} 
\ff
We also see that generally\footnote{The ratio of two 4-forms is a scalar which can be computed in any coordinate system.}
\f
\phi^{JK}  = \frac{1}{2g} \frac{  B^{(J} \wedge B^{K)}}{ B^L \wedge B_L}
\label{eq3.5}
\ff

Putting (\ref{eq1}) and (\ref{eq2}) together we find
\f
\left ( {\cal D} \phi_{JK} \right ) \wedge B^K =
 2g \phi^{KL} {\cal D}\phi_{KL} \wedge B_J
\label{eq4}
\ff

So far all equations are fully $G$ invariant.  I will now exhibit a natural
symmetry breaking mechanism that distinguishes an $SO(4)$ subalgebra,
leading to a coupling of general relativity to Yang-Mills fields in the
quotient $G/SO(4)$.

To exhibit the symmetry breaking mechanism we first 
decompose $G$ into its $SO(4)$ subalgebra and remaining 
generators in $G/SO(4)$.      As before, we  label the generators of $SO(4)$
by antisymmetric pairs of four dimensional vector indices, $[ab]$,
where $a=1,...4$.  The remaining generators in 
$G/SO(4)$ we label by
$i=7,8...n$.  Then we have $I= \{ [ab],,i \}$.  

We  look for a solution to the equations of motion to leading order in $g$ which breaks the symmetry from $G$ down to $SO(4)$.
To find it, we consider the component of (\ref{eq3}) in the $SO(4)$ 
directions
\f
B^{ab} \wedge B^{cd} = 2g  \phi^{abcd}B^L \wedge B_L
\label{ansatz1}
\ff
We require a solution to this to leading order in $g$ for which $B^L \wedge B_L$ is non-vanishing.  
This can be solved by taking as an ansatz a form discussed in  \cite{CMPR}, 
\f
\phi^{abcd}= \frac{1}{2g} \epsilon^{abcd} \ W+\frac{\rho}{2g} \delta^{a[c} \delta^{d]b} 
\label{ansatz2}
\ff
with the rest of the components of $\phi^{IJ}$ of order zero
in $g$ and higher.  Here $W$ is a scalar dimensionless function of the fields.  This breaks the symmetry, as it gives, 
to zeroth order in $g$, 
\f
\frac{B^{ab} \wedge B_{cd}}{B^L \wedge B_L} =  \epsilon^{ab}_{\ \ \ cd}  \  W 
+ \rho \delta^{[a}_c \delta^{b]}_c
\label{wow}
\ff

Tracing (\ref{ansatz1}) we find that
\f
\phi^{ab}_{\ \ \ ab} = \frac{6 \rho}{g} = \frac{1}{2g} \frac{B^{ab} \wedge B_{ab}}{B^L \wedge B_L}
\ff
It is consistent to assume that $B^i$ are of higher order in $g$ so that to leading order
\f
B^L \wedge B_L=B^{ab} \wedge B_{ab} + O(g)
\ff
which implies that
\f
\rho = \frac{1}{12} + O(g)
\ff

We next solve the remaining  components of  (\ref{eq3}) to order zero in $g$.  
To solve (\ref{ansatz1}) by ansatz (\ref{ansatz2}) we write, to leading order in $g$,
\f
B^{ab} =  \Sigma^{ab} + \gamma \Sigma^{* \ ab}
\label{Bis}
\ff
were $\Sigma^{ab} = e^a \wedge e^b$.  It follows that
\f
B^{ab}\wedge B_{ab}= 24\gamma e 
\ff
where $e$ is the determinant of the frame field.  $\gamma$ is a parameter which labels the solutions (\ref{Bis}), which is arbitrary but must not vanish.

We multiply (\ref{wow}) and (\ref{ansatz1}) by $\epsilon_{abcd}$ which tells us that
\f
W=\frac{1+2 \gamma^2}{24\gamma}
\ff

We now consider the equations satisfied by $B^i$.  
To zeroth order in $g$ we  have
have 
\f
B^{ab} \wedge B^i  =0
\ff
which becomes for solutions
\f
\Sigma^{ab} \wedge B^i  =0
\label{makevanish}
\ff
For each $i \in G/SO(4)$ these are six linear equations in the six components
of $B^i$ and they imply, so long as the metric is non-degenerate, 
\f
B^i=0
\ff
to zeroth order in $g$.

Thus we see that the ansatz (\ref{ansatz2}) has broken the symmetry.  The $n$  $B^I$ all start out on the same footing representing the full $G$ gauge symmetry.  But if $n > 6$ the
equations (\ref{eq3}) to zeroth order in $g$ 
\f
B^{(J} \wedge B^{K)} =0
\label{vanish}
\ff
cannot be solved to make more than $6$ of the $B^I$ non-vanishing.  This is bound to happen at some $n$ because there are $6n$ components of $B^I$ but $n(n+1)/2$ equations in (\ref{vanish}).  But we see that as soon as there are 
six non-vanishing $B^I$ valued in the Lorentz algebra, each additional 
$B^i$ must satisfy the $6$ linear equations (\ref{makevanish}).  These are six linear equations for the six components of each $B^i$, so the result is that
they all vanish to zeroth order in $g$.  They will be non-vanishing at 
$O(g)$.  Hence this is a mechanism to break $G$ down to the lorentz algebra
at a scale given by $g$.  

To higher order in $g$ we will  have
\begin{eqnarray}
B^{ab} &=&  \Sigma^{ab} + \gamma \Sigma^{* \ ab} + g b^{ab} + ...   \\
B^i &=& g b^i
\label{SIgma}
\end{eqnarray}
We will not here develop the solutions to higher order but it is straightforward
to do so. What we have so far is enough to see that 
we get the fields for general relativity coupled in some non-trivial way to additional matter fields represented by the fields with indices wholy or partly in $G/SO(4)$.  There will also be corrections to the Einstein and Yang-Mils equations given by a power series in $g$.  We note that the terms proportional to $g$, quadratic in $\phi_{IK}$ are necessary because otherwise the only solutions (\ref{eq3}) would have would lead to  most of the fields vanishing.

To see how the coupled Einstein-Yang-Mills dynamics arises, we can go use the equation of motion for 
$\phi^{IJ}$, equation (\ref{eq3}) and use it to eliminate the $\phi^{IJ}$.  This is fair to do since we have so far just used the consequences of the equations of motion for the $\phi_{IJ}$.  
Since in our ansatz $B^K \wedge B_K$ is non-vanishing, we use  (\ref{eq3.5})  to eliminate $\phi_{IJ}$.  
The result is  an
action equivalent to (\ref{action1}) on solutions to  (\ref{eq3}) where $B^K \wedge B_K$ is non-vanishing:
\f
S^{no \phi} = \frac{1}{G} \int B^I \wedge F_I 
- \frac{1}{8g} \frac{(B^I \wedge B^J )(B_I \wedge B_J )}{B^K \wedge B_K}
\label{wow2}
\ff
We can divide this into a gravity and matter part
\f
S^{no \phi} =S^{no \phi}_{grav}  +S^{no \phi}_{matter}
\ff
The gravity part is 
\begin{eqnarray}
S^{no \phi}_{grav} &=& \frac{1}{G} \int B^{ab} \wedge F_{ab}
- \frac{1}{8g} \frac{(B^{ab} \wedge B^{cd} )(B_{ab} \wedge B_{cd} )}{B^{ef} \wedge B_{ef}}
\nonumber \\
&=& \frac{1}{G_N} \int \left (  \Sigma ^{ab} \wedge F_{ab}^* + \gamma^{-1}  \Sigma ^{ab} \wedge F_{ab} 
-\frac{\Lambda }{2} e
\right )
\label{wow3}
\end{eqnarray}
where 
\f
\Lambda = \frac{1}{4g} \left [
1 +4 \gamma^2 + \gamma^4 
\right ]
\ff
and the Newton's constant is $G_N=G / \gamma$. 
Thus, we see that we recover the Palatini action with the Holst term, where $\gamma$ is the Immirzi parameter.  So our ansatz leads to standard gravity with a parity breaking topological term.  

To work out the matter part it is convenient to define
\f
F^i = \frac{1}{2} F^i_{ab} \Sigma^{ab}, \ \ \ \ \ B^i = \frac{1}{2}  B^i_{ab} \Sigma^{ab}
\ff
where $F^i_{ab}$ and $B^i_{ab}$ are functions.  Then we can expand the matter part of (\ref{wow2}) as
\f
S^{no \phi}_{matter} = \frac{1}{G} \int e \left (
\frac{1}{4}\epsilon^{abcd} B^i_{ab}F^i_{cd} -\frac{W }{64g} B_{iab}B^{iab} -\frac{1 }{24\cdot 64 g} 
\epsilon^{abcd} B^i_{ab}B^i_{cd} 
\right )
\ff
We can solve the equations of motion for $B^i_{ab}$ to find that 
\f
B^i_{ab}= g \xi  \left [  F^i_{ab} - 6W  F^{* i}_{ab}
\right ]
\ff
where $\xi =\frac{96}{1-36 W^2}$.

Plugging the solution into the action we find that
\f
S^{no \phi}_{matter} =-  \frac{1}{4} \int  \left ( \frac{e}{4 g_{YM}^2} F^{iab } F_{iab } 
+\theta  F^i \wedge F_i 
\right )
\label{YMlike}
\ff
where
\f
\frac{1}{g_{YM}^2}= \frac{W\xi}{ G_N\Lambda \gamma} (1+4 \gamma^2 +\gamma^4 )
 \left (
3 -\frac{\xi}{48}+\frac{3 }{8} \xi W^2
\right )
\ff
and
\f
\theta = \frac{-\xi}{16 G_N \Lambda \gamma} (1+4 \gamma^2 +\gamma^4 ) \left (
1 +\frac{\xi}{96}+\frac{9}{8} \xi W^2 
\right )
\ff

We see that roughly $g^2_{YM} \approx G_N \Lambda $   so the Yang-Mills theory is weakly coupled when the cosmological constant is small. It is also interesting to note that our ansatz naturally leads to a parity breaking topological term in the Yang-Mills sector, as it does for the gravitational part of the action.  However the $\theta$ angle is large.  

It is interesting to note that we get the Yang-Mills form (\ref{YMlike}) for the dynamics of all the components of the connection,
$A^i$ with components in $G/SO(4)$.  These include Yang-Mills fields in the largest compact subgroup of the quotient $G/SO(4)$, which we can call $H$. However, note that $F^i$ are the components of the full $G$ curvatures, which for $i \in H$ are the Yang-mills field strength plus corrections.   Note also that there are generally 
connection and curvature components in $G/(SO(4) \oplus H )$ which will not be Yang-Mills fields as they will transform nontrivially under both $H$ and $SO(4)$.  The details of this will be studied in \cite{SGL} for the case of $G=SO(8)$.  We will  also see there that the degrees of freedom in the off diagonal connections contain the Higgs fields for $H$.  

The scheme we have described works for any gauge group so it will
work for $G=E8$.  This gives a fully $E8$ covariant dynamics for Lisi's unification proposal\cite{lisi}.

\section{Discussion}

\subsection{Fermions}

\begin{figure}
\begin{center}
\includegraphics[height=2.5cm]{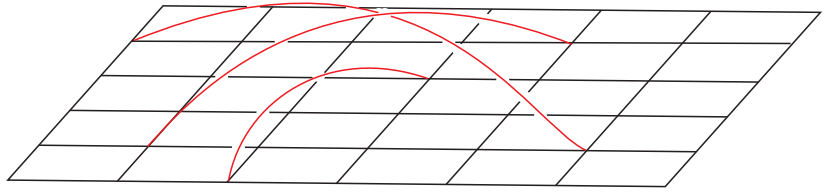}
\end{center}
\caption{Disordered locality: A regular graph contaminated by non-local links.}
\label{nll}
\end{figure}

In Lisi's proposal the fermions of the standard model are described by a $BRST$ extension of the connection.  There is within loop quantum gravity another alternative, which is that
fermions and scalars arise due to the possibility of disordered locality.  As described in \cite{disordered}
this phenomena arises because there are two notions of locality in loop quantum gravity.
Microscopic locality is determined by the connections of the spin network graphs, two
nodes are neighbors if they are connected.  In the semiclassical limit a classical metric will
emerge, generally from a superposition of spin network states.  This gives rise to a notion
of macrolocality.  As described in \cite{disordered,disordered2} these two notions of locality disagree when two nodes
$x$ and $y$, 
which are far from each other in terms of the classical emergent metric, are connected by a
single edge $e_{xy}$.  Furthermore, in those papers it is argued that
disordered locality is generically to be expected in the semiclassical limit. Consider a graph as in Figure (\ref{nll}) which is regular and therefor may occur in the superposition of states making up a semiclassical state associated with a flat metric.  There is in loop quantum gravity, no apparent energy cost to contaminating that lattice-like graph with non-local links as shown in the figure.  Nor is there an incompatibility with the semiclasicality of the state. As there are many more ways to add a link to a lattice that connects two far away nodes than two nearby nodes, there is an instability for the formation 
of such non-local links as the universe expands from Planck scales. Moreover, once inserted in a graph, non-local links are trapped, as they can only be eliminated if two of them annihilate by the coincidence of their ends arriving by local moves at neighboring nodes.  The proposal is then that these act as Planck scale wormholes, carrying quantum numbers associated with gauge fields carried by the non-local link.  

Let us consider observations made by a local observer in the neighborhood of $x$. From their point of view the edge $e_{xy}$ simply comes to an end, that is it appears to connect to a one valent node.   But ends, or one valent nodes in loop quantum gravity represent matter degrees
of freedom.  Thus, the dislocations due to disordered locality appear in the semiclassical limit as matter degrees of freedom.  

Let us suppose that the gauge group is  $SU(2) \otimes H$, where $H$ is an internal gauge symmetry. Then the edge $e_{xy}$ carries representations of these groups, $(j,r)$.  Local observers will describe  $e_{xy}$ as a particle of spin $j$ and charge $r$.  

This leads to a picture in which for every generator of $G$, the gauge symmetry, the
semiclassical limit has a gauge field plus a set of particle excitations given by the 
representations of $G$.  

This emergence of particle states is so far kinematical, more work needs to be done to ensure that in the low energy limit these particles have the correct dynamics and statistics.  Earlier work, described in \cite{me-fermions}, indicates that in the Hamiltonian constraint formulation of loop quantum gravity, Planck scale wormholes do behave as spinors.   

We then turn to some comments on the quantum dynamics.

\subsection{A class of spin foam models}

I would like to comment that the form of the extended Plebanski action may have a simple quantization in terms of a spin foam model.  Let us consider the action in the 
form (\ref{wow2}).  The Euclidean path integral naively will have the form,
\f
Z= \int dA dB e^{ \frac{1}{G} \int B^I \wedge F_I 
- \frac{\Lambda}{8G} \frac{(B^I \wedge B^J )(B_I \wedge B_J )}{B^K \wedge B_K}}
\ff
The $B\wedge F$ term alone would give rise to a topological field theory with a spin foam
formulation of the form of 
\f
Z= \sum_{r,i} \prod_{\mbox{4-simplices}}\{ \mbox{15-j } \}_G
\ff
where the four dimensional manifold has been triangulated whose faces are labeled by 
representations $r$ and the tetrahedra by the intertwiners $i$ of the group $G$. 
$\{ 15-j \}_G$ are the $15j$ symbols which are functions of the labels on each 
four-simplex.  The usual Barrett-Crane strategy is to modify this by constraining the
sums over representations and intertwiners to the balanced representations.  Instead, the
action (\ref{wow2}) suggests using the simplicity constraints as  Gaussian
weights so we have
\f
Z= \sum_{r,i} \prod_{\mbox{4-simplices}}\{  \mbox{15-j } \}_G e^{- \frac{\Lambda}{8G}
 \frac{(B^I \wedge B^J )(B_I \wedge B_J )}{B^K \wedge B_K}}
\ff
where the $B^I$ are functions of the representation labels on each face\footnote{A related form has been considered by Speziale for
$2+1$ gravity coupled to Yang-Mills field\cite{simone}.}. 

\subsection{Open issues}

I would like to close by listing a few out of many open issues facing this kind of unification.

\begin{itemize}

\item{} The kinematical quantum theory can now be developed along loop quantum gravity lines for a general $G$, as well as for the particular case of $E8$.

\item{} The spin foam quantization may also be explored based on the proposal discussed here.  It will be interesting to see  if the ultraviolet convergence results from the Barrett-Crane model also apply here. 

\item{}The proposal of matter as the ends of long distance links needs more development.  One needs to check whether the spin foam dynamics gives the right dynamics for the fermions in the case
of graviweak unification or a larger unification.  There are also open issues
regarding spin and statistics; these may be addressed by generalized
or topological spin-statistics theorems.

\end{itemize}

\section*{Acknowledgements}

I would like to thank Laurent Freidel, 
Sabine Hossenfelder, Jorge Pullin, Carlo Rovelli,  James Ryan and Eric Weinstein for helpful conversations, correspondence and suggestions. Thanks are also due to Garrett Lisi and Simone Speziale for very helpful discussions and comments, including finding an error in an earlier version of this paper.   Detailed comments from a referee were also very helpful for improving the presentation.  Research at
Perimeter Institute is supported in part by the Government
of Canada through NSERC and by the Province of Ontario through MEDT.

\end{document}